\documentstyle[12pt,epsfig]{article}  
\begin{document} 
\baselineskip=20pt

\def\la{\mathrel{\mathpalette\fun <}}
\def\ga{\mathrel{\mathpalette\fun >}}
\def\fun#1#2{\lower3.6pt\vbox{\baselineskip0pt\lineskip.9pt
\ialign{$\mathsurround=0pt#1\hfil##\hfil$\crcr#2\crcr\sim\crcr}}} 

\begin{titlepage} 
\begin{center}
{\Large \bf $P_T$-imbalance in dimuon+jet production as a 
signal of partonic energy loss in heavy ion collisions at LHC} \\ 

\vspace{4mm}

I.P.~Lokhtin,  A.V.~Sherstnev, and  A.M.~Snigirev,\\
M.V.Lomonosov Moscow State University, D.V.Skobeltsyn Institute of Nuclear 
Physics \\
119992, Vorobievy Gory, Moscow, Russia \\ E-mail:~~igor@lav01.sinp.msu.ru   
\end{center}  

\begin{abstract} 
We consider a hard jet production tagged by a muon pair in ultrarelativistic
heavy ion collisions. The process cross section is calculated by the CompHEP 
Monte-Carlo generator taking into account full $\gamma^*/Z$ interference pattern at LHC
energies. We have found that reasonable statistics, $\sim 1000$ events per 1 month of 
LHC run with lead beams, can be expected for realistic geometrical acceptance and 
kinematic cuts. The transverse momentum imbalance due to interactions of jet 
partons in the medium is evaluated for $\mu^+\mu^-$pair+jet correlation, as well as 
for the correlation between $\mu^+\mu^-$ pair and a leading particle in a jet.
Theoretical and experimental uncertainties of these observables are discussed.  
\end{abstract}

\bigskip

\noindent 
$PACS$: ~~12.38.Mh, 24.85.+p, 25.75.+r \\ 
$Keywords$:~~jet quenching, partonic energy loss, dileptons, 
relativistic nuclear collisions \\
\end{titlepage}   

\section{Introduction} 

One of the important tools to study properties of quark-gluon plasma (QGP)  in
ultrarelativistic heavy ion collisions is a QCD jet production. Medium-induced
energy loss of energetic partons, the so-called jet quenching, has been proposed to
be very different in cold nuclear matter and in QGP, resulting in many challenging
observable phenomena~\cite{baier_rev}. Recent RHIC data on suppression of inclusive
high-p$_T$ charge and neutral hadron production from STAR~\cite{star},
PHENIX~\cite{phenix}, PHOBOS~\cite{phobos} and BRAHMS~\cite{brahms} are in
agreement with the jet quenching hypothesis~\cite{Wang:2004}. However direct
event-by-event reconstruction of jets and their characteristics is not available in
RHIC experiments at the moment, while the assumption that integrated yield of all
high-$p_T$ particles originates only from jet fragmentation is not obvious.

At LHC a new regime of heavy ion physics will be reached at 
$\sqrt{s_{\rm NN}}=5.5$ TeV where hard and semi-hard QCD multi-particle production can dominate over
underlying soft events. The initial gluon densities in Pb$-$Pb reactions at LHC are
expected to be significantly higher than at RHIC, implying stronger partonic energy
loss which can be observable in various new
channels~\cite{lhc-jets,lhc-hq,lhc-phot}. In particular, the potentially important
process is production of a single jet opposite to a gauge boson in $\gamma
+$jet~\cite{Wang:1996} and ${\rm Z} +$jet~\cite{Kartvelishvili:1996} or a virtual
photon in $\gamma ^*(\rightarrow l^+l^-)+$jet~\cite{Awes:2003} final states,
dominantly through processes such as 
$${\rm q\:g}\rightarrow{\rm q}\:\gamma\:,\;\;\;\;\; {\rm qg\:}\rightarrow
{\rm q\:Z}\:,\;\;\;\;\;{\rm q\:g}\rightarrow{\rm q}\:\gamma^*\:.$$
In heavy ion collisions, the relative $p_T$ between the jet (or leading
particle in a jet)  and the boson becomes imbalanced due to interactions of the jet
partons in the medium.

In the $\gamma +$jet case the main problem arises from the jet pair production 
background when a leading $\pi^{0}$ in the jet is misidentified as a photon. The 
``photon isolation'' criteria usually used in $pp$ collisions do not work with the 
same efficiency in high multiplicity heavy ion interactions~\cite{Kodolova:2003}. 
Thus, the question of adequate using photon+jet correlation to study jet quenching
requires further investigation. On the other hand, the production of jet tagged by
dileptons is not affected significantly by backgrounds and can be used to observe
$p_T$-imbalance as a signal of medium-induced partonic energy loss.

In this Letter we analyze dimuon+jet production (including both 
$\gamma^*/Z\rightarrow\mu^+\mu^-$ modes) in heavy ion collisions at the LHC. 
In Sect.~2 the cross section of this process is calculated by CompHEP Monte-Carlo
generator and the expected event rate is estimated for realistic geometrical
acceptance and kinematic cuts. Sect.~3 describes shortly the model of partonic
energy loss in QGP used to calculate $p_T$-imbalance between $\mu^+\mu^-$ pair and
jet (or leading particle in a jet). Discussion on numerical results and their
experimental and theoretical uncertainties can be found in Sect.~4, summary - in 
Sect.~5. 

\section{Dilepton+jet production at LHC}

We use the CompHEP Monte-Carlo generator~\cite{comphep} for cross section 
calculation of dilepton+jet process and subsequent generation of event for this 
process. CompHEP is a general, tree level generator, which allows one to study 
almost all processes $2\rightarrow N$ (up to $N=5$) in the framework of the 
usual technique of Feynman diagrams squared for different models (SM, MSSM, many 
other...). It generates, squares and symbolically calculates a set of Feynman 
diagrams for a given process and creates a numeric Monte-Carlo generator for 
the process. This MC generator allows one to compute cross sections (with 
applied cuts), to build distributions and to generate events with partons in 
the final state; the initial partons are convoluted with parton distribution 
functions (PDF).

For simplification of our calculation and event generation we apply a special 
{\it hash}-model in CompHEP~\cite{Boos:2000ap}. In this model a unitary 
rotation of down quarks transfers Cabibbo-Kabayashi-Maskawa (CKM) matrix 
elements from interaction vertices to parton distribution functions. It allows 
one to unify two light quark 
generations to one only. As a result, this trick reduces significantly number 
of subprocesses which we need to take into account for the process.
This model applies two approximations: 
\begin{itemize}

\item 
$u_h$ ($u$, $c$) and $d_h$ ($d$, $s$) quarks are massless; 

\item 
$b$ and $t$ quarks do not interact with light quarks and CKM matrix element 
$V_{tb}$=1.
\end{itemize}
In the problem of dilepton+jet investigation the influence of factors violating  
these conditions is very small so we can soundly use this approximation.

There are nine subprocesses contributed to the process of the dilepton+jet production
in the framework of SM with {\it hash}-approximation. Feynman diagrams
for the subprocesses are depicted on fig.1. We use the following 
physics parameter values: $\alpha=1/127.9$, $\sin{\theta_W}=0.2311$, 
$M_Z=91.1876$ GeV, $\Gamma_Z=2.4368$ GeV, $M_{\mu}=105.7$ MeV, $M_b=4.85$ GeV, 
PDF are taken from cteq5l~\cite{Lai:1999wy}.
We apply a following set of ``loose'' cuts for event generation. In the further 
investigation these cuts have been strengthened:
\begin{itemize}

\item 
$p_T^{\mu}>5$ GeV/$c$  and $p_T^{\rm jet}>20$ GeV/$c$;   

\item 
$\mid \eta^{\mu,~\rm jet}\mid <3$.
\end{itemize}
We do not apply a cut on $M(\mu^+\mu^-)$ because of a singularity in the region
$M(\mu^+\mu^-)\rightarrow0$ is regularized by massiveness of muon. Contributions of
all subprocess to the total cross section is presented in the 
Table~\ref{crosssections}.

\begin{table*}
\begin{center}
\caption{Contributions to process $pp\rightarrow \mu^-\mu^++{\rm jet}$ 
($\sqrt{s_{\rm pp}}=5.5$ TeV)}

\bigskip 

\begin{tabular}{|c|c|}
\hline
Subprocess:                    & Cross section (pb):\\
\hline
$u_h\bar{u_h}\rightarrow \mu^-\mu^++g        $ &  27.5 \\
$d_h\bar{d_h}\rightarrow \mu^-\mu^++g        $ &  18.1 \\
$u_hg        \rightarrow \mu^-\mu^++u_h      $ & 105.0 \\
$d_hg        \rightarrow \mu^-\mu^++d_h      $ &  36.2 \\
$\bar{u_h}g  \rightarrow \mu^-\mu^++\bar{u_h}$ &  44.3 \\
$\bar{d_h}g  \rightarrow \mu^-\mu^++\bar{d_h}$ &  21.9 \\
$b\bar{b}    \rightarrow \mu^-\mu^++g        $ &   0.6 \\
$bg          \rightarrow \mu^-\mu^++b        $ &   3.6 \\
$\bar{b}g    \rightarrow \mu^-\mu^++\bar{b}  $ &   3.6 \\
\hline
$qq          \rightarrow \mu^-\mu^++q        $ & 260.9 \\
\hline
\end{tabular}\label{crosssections}
\end{center}
\end{table*}

\vskip -5mm

Note that the cross section for the process $\mu^+\mu^-$+jet was estimated in 
ref.~\cite{Awes:2003} at RHIC and LHC energies taking into account the
$\gamma^*$ contribution only. Such approximation seems valid for RHIC, but for 
LHC the contribution from $Z$ and $\gamma^*/Z$ interference term are also
significant. 

Now we estimate the expected event rate for realistic
geometrical acceptance and kinematic cuts. To be specific, the geometry of Compact
Muon Solenoid (CMS) detector is considered~\cite{cms94,note00-060}: pseudo-rapidity
coverage $\mid \eta \mid < 3$ for jets and $\mid \eta \mid < 2.4$ for muons. 
Extra cuts $P_T^{\mu^+\mu^-}> 50$ GeV/$c$ and $E_T^{\rm jet}>50$ GeV were 
applied. Then the corresponding $pp$ cross section for $\mu ^+\mu ^-$+jet
production is $\approx 16~p$b, and Pb$-$Pb cross section is estimated as 
$16~{\rm pb} \times (207)^2 \approx 0.7~\mu$b. The corresponding event rate in a one 
month Pb$-$Pb run (assuming 15 days of data taking), $T=1.3\times 10^6$ s, with 
designed luminosity $L = 10^{27}~$cm$^{-2}$s$^{-1}$, is $N_{\rm ev}= T 
\sigma_{PbPb} L \approx 1000$ in this case. Note that potential using also 
$e^+e^-$+jet channel could increase observed event rates by a factor $\sim 2$ 
and requires further investigation.

To conclude this section, let us discuss the potential background sources
in the CMS experimental situation. 
Semileptonic heavy quark decays and uncorrelated pion and kaon decays are expected 
to give main contributions to the dimuon spectra at LHC energies, $\sim 10^5$ 
events in a one month Pb$-$Pb run for CMS acceptance~\cite{lhc-hq,note00-060}. 
However, the request of hard enough cut on muon pair transverse momentum, 
$P_T^{\mu^+\mu^-}> 50$ GeV, together with the additional trigger to have a hard jet 
with $E_T^{\rm jet}>50$ GeV in the opposite hemisphere, makes such background 
``contamination'' negligible. Moreover, the experimental control on background 
extraction may be done by monitoring uncorrelated and correlated sources independently. 
The uncorrelated part can be subtracted using like-sign dimuon mass spectra, while 
the correlated part can be rejected using tracker information on the dimuon vertex 
position~\cite{lhc-hq,note00-060}.

\section{Simulation of jet quenching at LHC} 

In order to generate the initial distributions of jets and $\mu^ +\mu^ -$ pairs 
in nucleon-nucleon sub-collisions at $\sqrt{s}=5.5$ TeV, we have used CompHEP
package for initial parton configuration setting and 
PYTHIA$\_6.2$~\cite{pythia} for subsequent jet fragmentation.  
After specifying initial partonic state, event-by-event Monte-Carlo simulation of 
rescattering and energy loss of partons in QGP is performed (for details of the
model one can refer to~\cite{lokhtin98,lokhtin00}). 
The approach relies on accumulative energy losses, when gluon radiation is 
associated with each scattering in the expanding medium together with including 
the interference 
effect by the modified radiation spectrum $dE/dl$ as a function of decreasing 
temperature $T$. The basic kinetic integral equation for the energy loss $\Delta E$ 
as a function of initial energy $E$ and path length $L$ has the form 
\begin{eqnarray} 
\label{elos_kin}
\Delta E (L,E) = \int\limits_0^Ldl\frac{dP(l)}{dl}
\lambda(l)\frac{dE(l,E)}{dl} \, , ~~~~ 
\frac{dP(l)}{dl} = \frac{1}{\lambda(l)}\exp{\left( -l/\lambda(l)\right) }
\, ,  
\end{eqnarray} 
where $l$ is the current transverse coordinate of a parton, $dP/dl$ is the scattering
probability density, $dE/dl$ is the energy loss per unit length, $\lambda = 1/(\sigma 
\rho)$ is in-medium mean free path, $\rho \propto T^3$ is the medium density at 
the temperature $T$, $\sigma$ is the integral cross section of parton 
interaction in the medium.
Such a numerical simulation of the free path of a hard jet in QGP allows any 
kinematic characteristic distributions of jets in the final state to be obtained. 

The collisional energy loss due to elastic scattering with high-momentum transfer 
have originally been estimated by Bjorken in~\cite{bjork82}, and recalculated later 
in~\cite{mrow91} taking also into account the loss with low-momentum transfer
dominated by the interactions with plasma collective modes. Since the latter 
process contributes to the total collisional loss without the large factor 
$\sim \ln{(E / \mu_D)}$ ($\mu_D$ is the Debye screening mass) in comparison with 
high-momentum scattering and it can be effectively ``absorbed'' by the redefinition 
of minimal momentum transfer $t \sim \mu_D^2$ under the numerical estimates, 
we used the collisional part with high-momentum transfer only~\cite{lokhtin00},   
\begin{equation} 
\label{col} 
\frac{dE}{dl}^{col} = \frac{1}{4T \lambda \sigma} 
\int\limits_{\displaystyle\mu^2_D}^
{\displaystyle 3T E / 2}dt\frac{d\sigma }{dt}t ~,
\end{equation} 
and the dominant contribution to the differential cross section 
\begin{equation} 
\label{sigt} 
\frac{d\sigma }{dt} \cong C \frac{2\pi\alpha_s^2(t)}{t^2} ~,~~~~
\alpha_s = \frac{12\pi}{(33-2N_f)\ln{(t/\Lambda_{QCD}^2)}} \>
\end{equation} 
for scattering of a parton with energy $E$ off the ``thermal'' partons with energy 
(or effective mass) $m_0 \sim 3T \ll E$. Here $C = 9/4, 1, 4/9$ for $gg$, $gq$ and 
$qq$ scatterings respectively, $\alpha_s$ is the QCD running coupling constant for 
$N_f$ active quark flavors, and 
$\Lambda_{QCD}$ is the QCD scale parameter which is of the order of the critical 
temperature,  $\Lambda_{QCD}\simeq T_c \simeq 200$ MeV. The integrated cross 
section $\sigma$ is regularized by the Debye screening mass squared $\mu_D^2
(T) \simeq 4\pi \alpha _s T^2(1+N_f/6)$. 

There are several calculations of the inclusive energy distribution of
medium-induced gluon radiation from Feyman multiple scattering diagrams. The 
relation between these approaches and their main parameters were discussed in
details in the recent writeup of the working group ``Jet Physics'' for the 
CERN Yellow Report~\cite{lhc-jets}. We restrict to ourself here by using BDMS
formalism~\cite{baier}. In the BDMS framework the strength of multiple
scattering is characterized by the transport coefficient 
$\hat{q}=\mu_D^2/\lambda_g $ ($\lambda_g$ is the gluon mean free path), which is
related to the elastic scattering cross section $\sigma$ (\ref{sigt}). In our
simulations this strength in fact is regulated mainly by the initial  
temperature $T_0$. Then the energy spectrum of coherent medium-induced gluon 
radiation and the corresponding dominated part of the radiative energy loss has 
the form~\cite{baier}: 
\begin{eqnarray} 
\label{radiat} 
\frac{dE}{dl}^{rad} = \frac{2 \alpha_s (\mu_D^2) C_R}{\pi L}
\int\limits_{\omega_{\min}}^E  
d \omega \left[ 1 - y + \frac{y^2}{2} \right] 
\>\ln{\left| \cos{(\omega_1\tau_1)} \right|} 
\>, \\  
\omega_1 = \sqrt{i \left( 1 - y + \frac{C_R}{3}y^2 \right)   
\bar{\kappa}\ln{\frac{16}{\bar{\kappa}}}}
\quad \mbox{with}\quad 
\bar{\kappa} = \frac{\mu_D^2\lambda_g  }{\omega(1-y)} ~, 
\end{eqnarray} 
where $\tau_1=L/(2\lambda_g)$, $y=\omega/E$ is the fraction of the hard parton 
energy carried by the radiated gluon, and $C_R = 4/3$ is the quark color factor. 
A similar expression for the gluon jet can be obtained by substituting 
$C_R=3$ and a proper change of the factor in the square bracket in (\ref{radiat}), see
ref.~\cite{baier}. The integral (\ref{radiat}) is carried out over all energies from 
$\omega_{\min}=E_{LPM}=\mu_D^2\lambda_g$, the minimal radiated gluon energy in 
the coherent LPM regime, up to initial jet energy $E$. 
 
The medium was treated as a boost-invariant longitudinally expanding quark-gluon 
fluid, and partons as being produced on a hyper-surface of equal proper times 
$\tau$~\cite{bjorken}. In order to simplify numerical calculations (and not to 
introduce new parameters) we omit the transverse expansion and viscosity of the fluid 
using the well-known scaling solution due to Bjorken~\cite{bjorken} for a 
temperature and density of QGP at $T > T_c \simeq 200$ MeV:
\begin{equation}
\varepsilon(\tau) \tau^{4/3} = \varepsilon_0 \tau_0^{4/3},~~~~
T(\tau) \tau^{1/3} = T_0 \tau_0^{1/3},~~~~ \rho(\tau) \tau = \rho_0 \tau_0 .
\end{equation}
Let us remark that the influence of the transverse flow, as well as of the mixed phase
at $T = T_c$, on the intensity of jet rescattering (which is a strongly increasing
function of $T$) seems to be inessential for high initial temperatures
$T_0 \gg T_c$~\cite{lokhtin98,lokhtin00}. On the contrary, the presence of viscosity 
slows down the cooling rate, which leads to a jet parton spending more time in the 
hottest regions of the medium. As a result the rescattering intensity goes up, i.e., in 
fact the effective temperature of the medium is increased as compared with the perfect 
QGP~\cite{lokhtin98,lokhtin00}. We also do not take into account here the 
probability of jet rescattering in nuclear matter, because the intensity of this process 
and the corresponding contribution to the total energy loss are negligible due 
to the much smaller energy density in ``cold'' nuclei. For certainty we used the initial 
conditions for the gluon-dominated plasma formation expected for central Pb$-$Pb 
collisions at LHC~\cite{esk}: 
$\tau_0 \simeq 0.1~{\rm fm/c}, ~T_0 \simeq 1~{\rm GeV}, ~\rho_g 
\approx 1.95T^3$.  
For non-central collisions we suggest proportionality of the initial energy 
density $\varepsilon _0$ to the ratio of the nuclear overlap function and the 
effective transverse area of nuclear overlapping~\cite{lokhtin00}.

In each event the distribution of jet production vertex at the given impact parameter 
$b$ of $AA$ collision is generated according to the distribution~\cite{lokhtin00}  
\begin{equation}
\label{vertex}
\frac{dN^{\rm jet}}{d\psi dr} (b) = \frac{T_A(r_1) T_A(r_2)}
{\int\limits_0^{2\pi} d \psi \int\limits_0^{r_{max}}r dr T_A(r_1) T_A(r_2)} , 
\end{equation} 
where $r_{1,2} (b,r,\psi)$ are the distances between the nucleus centers and the jet
production vertex $V(r\cos{\psi}, r\sin{\psi})$; $r_{max} (b, \psi) \le R_A$ is the 
maximum possible transverse distance $r$ from the nuclear collision axis to the $V$; 
$R_A$ is the radius of the nucleus $A$; $T_A(r_{1,2})$ is the nuclear thickness 
function (see ref.~\cite{lokhtin00} for detailed nuclear geometry explanations). 
After that, in every $i$th scattering of the co-moving particle (with the same 
longitudinal rapidity) a fast parton loses energy in the collisions and radiatively, 
$\Delta e_i = t_i/(2m_0) + \omega _i$, where $t_i$ and $\omega _i$ are simulated 
according to eqs.~(\ref{col}) and (\ref{radiat}) respectively. Thus in each event 
the energy of an initial parton decreases by the value 
$\Delta E (r, \psi )= \sum _i \Delta e_i$. 

In the frame of this model and using above QGP parameters we evaluate the mean energy 
loss of quark of $E_T=50$ GeV in minimum-bias Pb$-$Pb collisions, 
$\left< \Delta E_T^q \right> \sim 25$ GeV. In order to analyze the sensitivity of 
dimuon-jet correlations to the absolute value of partonic energy loss, we also performed 
the same calculations for the reduced initial temperature, $T_0=0.7$ GeV, which results 
in decreasing average energy loss by a factor $\sim (1/0.7)^3 \approx 3$. 

The distribution over a difference between $P_T^{\mu ^+\mu ^-}$, transverse momentum of 
$\mu ^+\mu ^-$ pair, and $E_T^{\rm jet}$, observed transverse energy of jet,
depends crucially on a fraction of the partonic energy loss falling {\it outside} 
the jet cone. There are some discussions on the angular spectrum of in-medium radiated 
gluons in the literature~\cite{lokhtin98,baier,Zakharov:1999,urs,vitev}. 
In fact, since coherent Landau-Pomeranchuk-Migdal radiation induces a strong 
dependence of the radiative energy loss of a jet on the angular cone size, it will 
soften particle energy distributions inside the jet, increase the multiplicity of 
secondary particles, and to a lesser degree, affects the total jet energy. On the 
other hand, the energy loss in the collisions turns out to be practically independent 
of the jet cone size and causes the loss of the total jet energy, because the bulk of 
the ``thermal'' particles knocked out of the dense matter by the elastic scatterings flies 
away in on almost transverse direction relatively to the jet axis~\cite{lokhtin98}. 
Thus, although the radiative energy loss of an 
energetic parton dominates over the loss in the collisions by up to an order of 
magnitude, the relative contribution of the latter jet energy loss grows with 
increasing the jet cone size due to the essentially different angular structure of loss 
for two mechanisms~\cite{lokhtin98}. Moreover, the total energy loss of a jet 
will be sensitive to the experimental capabilities to detect low-p$_T$ 
particles -- products of soft gluon fragmentation: thresholds for a giving signal 
in calorimeters, influence of the strong magnetic field, etc.~\cite{note00-060}. 

Since the full treatment of the angular spectrum of emitted gluons is rather 
sophisticated and 
model-dependent~\cite{lokhtin98,baier,Zakharov:1999,urs,vitev}, we considered 
two simple parameterizations of the distribution of in-medium radiated gluons over
the emission angle $\theta$. The ``small-angular'' radiation spectrum was
parameterized in the form
\begin{equation} 
\label{sar} 
\frac{dN^g}{d\theta}\propto \sin{\theta} \exp{\left( -\frac{(\theta-\theta
_0)^2}{2\theta_0^2}\right) }~, 
\end{equation}
where $\theta_0 \sim 5^0$ is the typical angle of the coherent gluon radiation estimated
in~\cite{lokhtin98}. The "broad-angular" spectrum has the form 
\begin{equation} 
\label{war} 
\frac{dN^g}{d\theta}\propto \frac{1}{\theta}~.  
\end{equation}
We believe that such a simplified treatment here is enough to demonstrate
the sensitivity of $p_T$-imbalance in $\mu ^+\mu ^-$+jet production to
the medium-induced partonic energy loss. 

\section{Numerical results and discussion} 

To be specific, the jet energy is defined here as the total transverse energy 
of the final particles collected around the direction of a leading particle inside 
the cone $R=\sqrt{\Delta \eta ^2+\Delta \varphi ^2}=0.5$, where $\eta$ and $\varphi$ 
are the pseudorapidity and the azimuthal angle respectively. Fig.2 shows the 
distribution over $\left( P_T^{\mu ^+\mu ^-}-E_T^{\rm jet}\right) $ for the cases 
without and with medium-induced energy loss obtained in the framework of our
model in minimum-bias Pb$-$Pb collisions, two parameterizations of distribution
on gluon emission angles (\ref{sar}) and (\ref{war}) being used. The same
geometrical acceptance and kinematic cuts as in Sect.2 were applied: 
$\mid \eta^{\rm jet} \mid < 3$, $\mid \eta^{\mu} \mid < 2.4$, 
$p_T^{\mu} > 5$ GeV/$c$, $P_T^{\mu ^+\mu ^-}, E_T^{\rm jet}>50$ GeV. 
Although $\left( P_T^{\mu ^+\mu ^-}-E_T^{\rm jet}\right) $-distribution is 
already smeared in 
$pp$ case, mainly due to the initial state gluon radiation, the mean value 
$\left< P_T^{\mu ^+\mu ^-}-E_T^{\rm jet}\right> $, as well as the maximum of the
distribution are close to $0$. The partonic energy loss in heavy ion collisions
results in the visible asymmetry of this distribution, its smearing and
shifting mean and maximum values. The effect is more pronounced for the 
``broad-angular'' radiation, because the contribution of the ``out-of-cone'' 
partonic energy loss is larger as compared with the ``small-angular'' radiation 
case. It is important to note that the contribution to jet-dimuon 
$P_T$-imbalance from collisional part (always ``out-of-cone'') is rather 
dominant over the contributions from ``broad-angular'' and ``small-angular'' 
radiation. The latter (dash-dotted 
histrogram) does not disappear totally just due to the fact that not only 
leading (parent) parton, but all partons of a jet pass through the dense 
medium and emit gluons under the angles $\theta$ relatively to their proper 
directions, which in general may not coincide with the jet axis (determined by 
the direction of a leading particle) and sometimes be even at the jet periphery.
Thus dimuon-jet correlation will be affected strongly by the collisional 
part of the loss.   

In the real experimental situation the jet observables will be sensitive 
to the accuracy of jet energy reconstruction in a high multiplicity environment. 
There are the two terms determining this accuracy: systematic jet energy loss 
and jet energy resolution (due to calorimeter and jet finder peculiarities, 
influence of magnetic field, etc.). Short measure of jet energy will be well-controlled  
systematic error which has the similar values for heavy ion and for $pp$ 
collisions, and it can be taken into account using the standard calibration 
procedure~\cite{note02-023}. On the 
other hand, the finite jet energy resolution will result in additional 
smearing $\left( P_T^{\mu ^+\mu ^-}-E_T^{\rm jet}\right) $-distribution without 
shifting mean and maximum values. In order to illustrate the sensitivity 
of such observables to experimental jet energy resolution $\sigma_{E_T}$, we set for 
definiteness the simple parameterization, $\sigma_{E_T}=1.5 \sqrt{E_T}$ GeV, 
which is close to one obtained with window-type jet finding algorithm for 
central Pb-Pb collisions (with background generated by fast MC code 
HYDRO~\cite{hydro} for multpilicity $dN^{\pm }/dy (y=0)=8000$) at 
CMS~\cite{lhc-jets}. Fig.3 shows the 
same distribution as fig.2, but final jet energy was smeared in each event by 
value $\sigma_{E_T}$. Due to additional smearing the initial distribution, the 
effect of jet quenching on shifting mean and maximum values becomes less 
visible and rather marginal for moderate partonic energy loss. 
  
Since jet observables are affected by a number of theoretical 
(in particular, sensitivity to the angular spectrum of medium-induced radiation and 
to the collisional part of energy loss) 
and methodical (finite jet energy resolution) uncertainties, complementary 
leading particle measurements are potentially important, such as $p_T$-imbalance between 
muon pair and a leading particle in a jet. Fig.4 presents the distribution over 
variable $\left( P_T^{\mu ^+\mu ^-} - 5\times E_T^{\rm leader}\right) $ 
(under the same conditions as for fig.2), where $E_T^{\rm leader}$ is the 
transverse energy of a leading particle (i.e. particle with maximum $E_T$) in a 
jet. The choice of a factor of $5$ denotes only the fact that the most 
probable value of a fraction of jet energy carried by leading particles is 
$\sim 0.2$ in our case. This distribution is originally more smeared and asymmetric 
than the distribution over $\left( P_T^{\mu ^+\mu ^-}-E_T^{\rm jet}\right) $. 
However additional smearing and shifting mean and maximum values due to 
partonic energy loss can be also clearly seen even for relatively small loss, 
$\left< \Delta E_T^q \right> \sim 8$ GeV at $T_0=0.7~(b=0)$ GeV. Moreover, the 
observed $p_T$-imbalance between $\mu ^+\mu ^-$ pair and a leading particle in 
a jet is directly related to the absolute value of partonic energy loss and almost 
insensitive to the form of the angular spectrum of emitted gluons and
collisional loss. The small 
difference for various angular distributions is just due to the moderate 
distinction of event samples which were triggered by having a jet with 
$E_T^{\rm jet}>50$ GeV. Since fixing the minimum threshold for jet detection is
sensitive to the total jet energy (and consequently to the angular dependence of
energy loss), the small influence of the latter on dimuon-leader correlation appears. 
We also become convinced that taking into account the jet energy smearing 
$\sigma_{E_T}$ almost does not have influence on dimuon-leader correlation
(see fig.5).  

\section{Conclusions} 
 
The channel with dimuon tagged jet production in ultrarelativistic 
heavy ion collisions was analyzed. The cross section of this process and 
corresponding event rates at LHC energies were evaluated with CompHEP
Monte-Carlo generator taking into account full $\gamma^*/Z$ interference pattern. 
The reasonable statistics, $\sim 1000$ events per 1 month of LHC run with lead 
beams, can be expected for realistic geometrical acceptance and kinematic 
cuts.

The correlations between $\mu ^+\mu ^-$ pair and jet, as well as between 
$\mu ^+\mu ^-$ pair and a leading particle in a jet, were first numerically 
studied for heavy ion collisions. The medium-induced partonic energy loss can 
result in significant smearing the distribution on difference between the 
transverse momentum of $\mu ^+\mu ^-$ pair and the jet transverse energy, and  
shifting mean and maximum values of the distribution. This effect will be
sensitive to the fraction of partonic energy loss (dependent on the form of 
the angular spectrum of in-medium radiated gluons) falling outside the jet 
cone. However the finite experimental jet energy resolution can result in 
additional smearing dimuon-jet correlation, which can make difficult 
observation of $P_T$-imbalance especially for moderate partonic energy loss. 

Since jet observables are affected by a number of theoretical and methodical 
uncertainties, complementary leading particle measurements will be useful and
even preferable. $P_T$-imbalance between $\mu ^+\mu ^-$ pair and a leading 
particle in a jet is quite visible even for moderate loss, directly related to  
the absolute value of partonic energy loss, almost insensitive to the angular 
spectrum of emitted gluons and to experimental jet energy resolution. 

Finally, the study of $\mu^+\mu^-$+jet and $\mu^+\mu^-$-leading-particle 
correlations is important for extracting the information about medium-induced 
partonic energy loss and properties of super-dense QCD matter to be created in 
heavy ion collisions at the LHC. 

{\it Acknowledgments.}  
Discussions with E.E.~Boos and L.I.~Sarycheva are gratefully acknowledged. 
This work is supported by grants N 04-02-16333 and N 04-02-17448 of Russian 
Foundation for Basic Research, and the grant of the Program ``Universities of 
Russia'' UR.02.03.028.

\begin{figure} 
\begin{center} 
\makebox{\epsfig{file=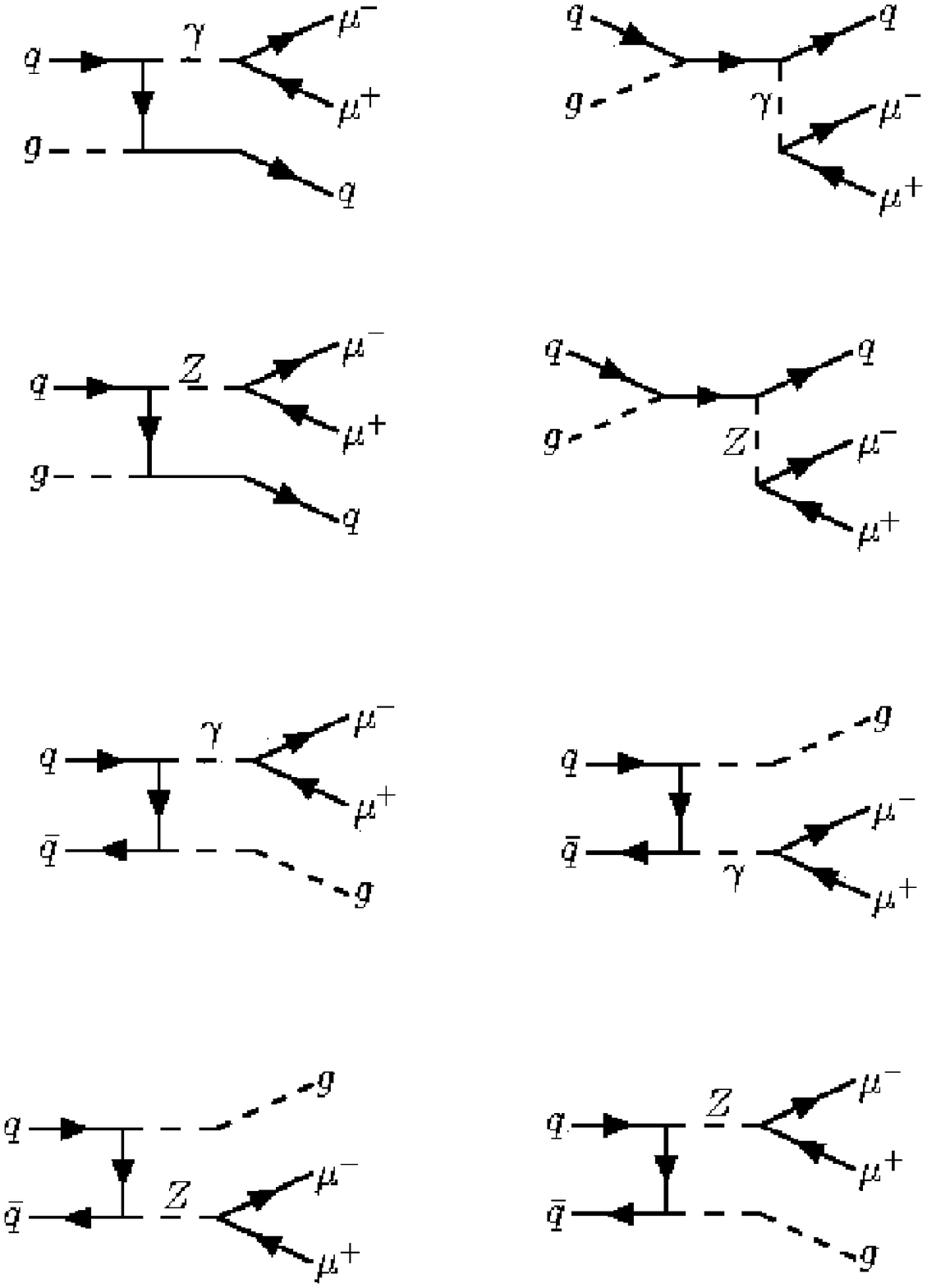, height=180mm}}   
\caption{Feynman diagrams for processes $qg\rightarrow \mu^-\mu^++q$ and 
$q\bar{q}\rightarrow \mu^-\mu^++g$ in the {\it hash} approximation of SM.} 
\end{center}
\end{figure}

\begin{figure} 
\begin{center} 
\makebox{\epsfig{file=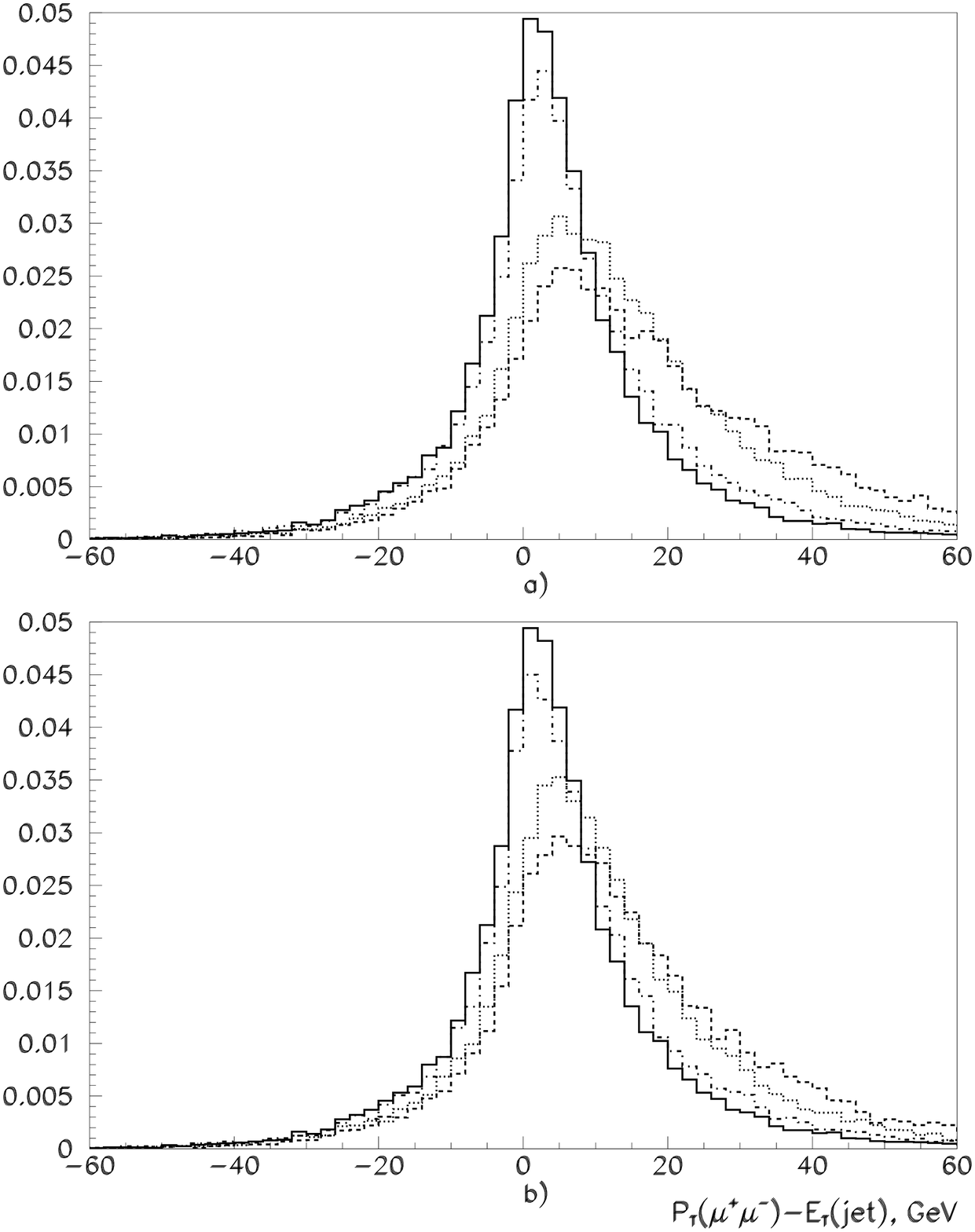, height=180mm}}   
\vskip -5 mm
\caption{The distribution over difference between the transverse momentum of 
$\mu ^+\mu ^-$ pair, $P_T^{\mu ^+\mu ^-}$, and the jet transverse energy, 
$E_T^{\rm jet}$, without (solid histogram) and with medium-induced partonic 
energy loss for the ``small-angular'' (\ref{sar}) (dotted histogram - radiaitve
and collisional loss, dash-dotted historgam - radiaitve loss only) and the 
``broad-angular'' (\ref{war}) (dashed histogram) parameterizations of emitted 
gluon spectrum in minimum-bias Pb$-$Pb collisions. Applied kinematical cuts are
described in the text. Initial QGP temperature $T_0=1~(b=0)$ GeV (a) and 
$T_0=0.7~(b=0)$ GeV (b).} 
\end{center}
\end{figure}

\begin{figure} 
\begin{center} 
\makebox{\epsfig{file=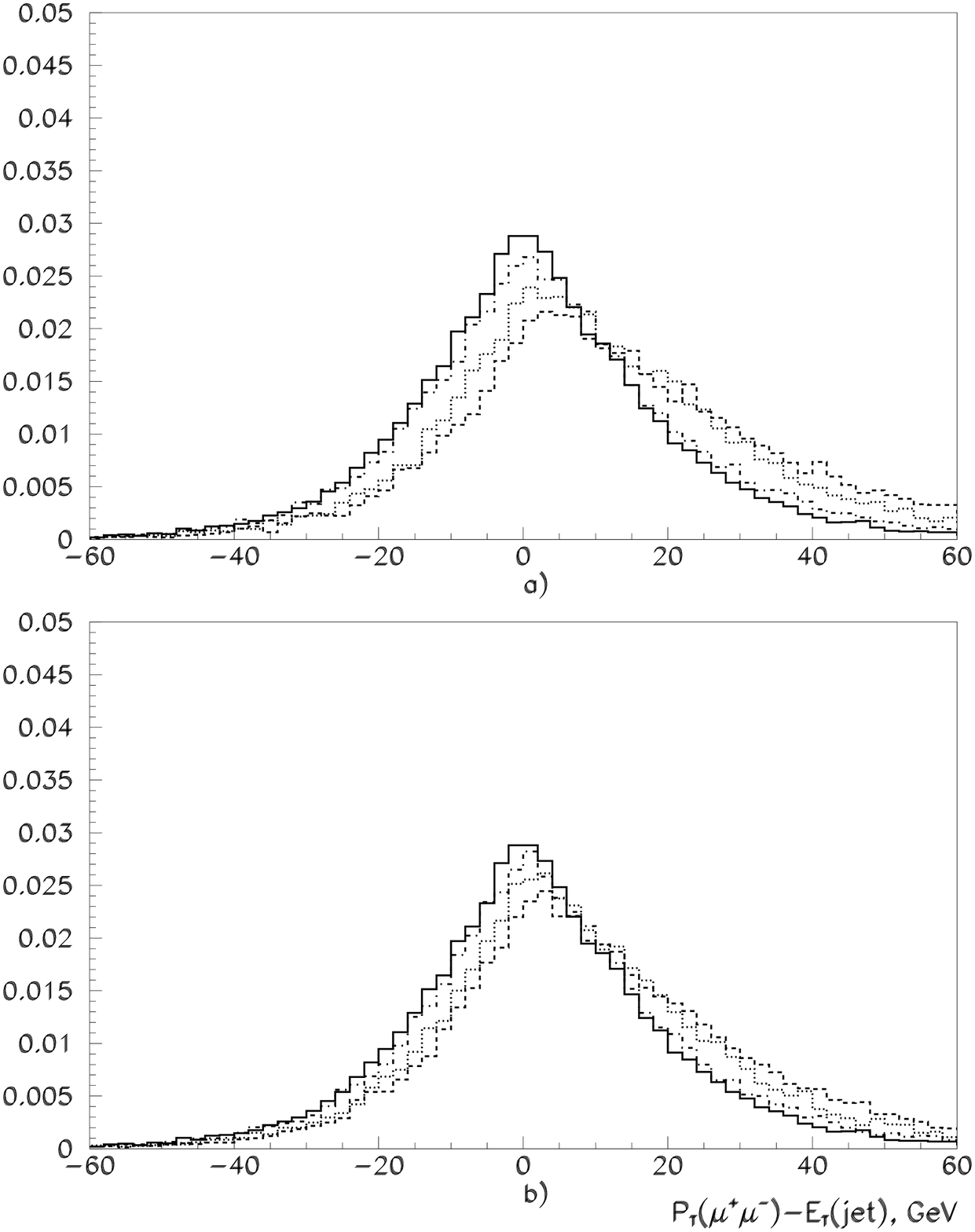, height=180mm}}   
\vskip -5 mm
\caption{The same as in fig.2, but including jet energy ``smearing'',
$\sigma_{E_T}=1.5 \sqrt{E_T}$ GeV.} 
\end{center}
\end{figure}

\begin{figure} 
\begin{center} 
\makebox{\epsfig{file=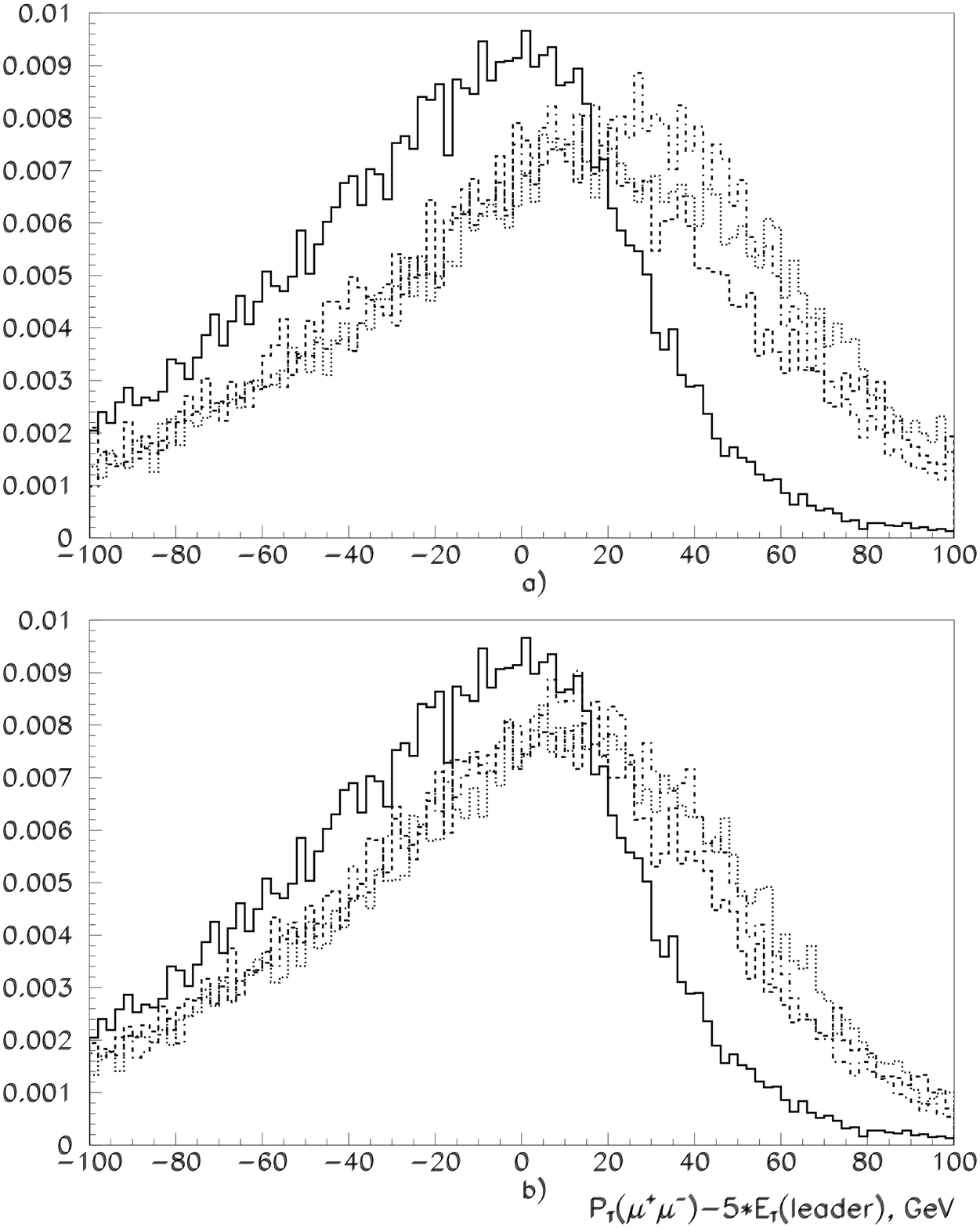, height=180mm}}
\vskip -5 mm    
\caption{The distribution over difference between the transverse momentum of 
$\mu ^+\mu ^-$ pair, $P_T^{\mu ^+\mu ^-}$, and five times transverse energy of 
a leading particle in a jet, $5 \times E_T^{\rm leader}$. The other conditions 
are the same as for fig.2.}
\end{center}
\end{figure}

\begin{figure} 
\begin{center} 
\makebox{\epsfig{file=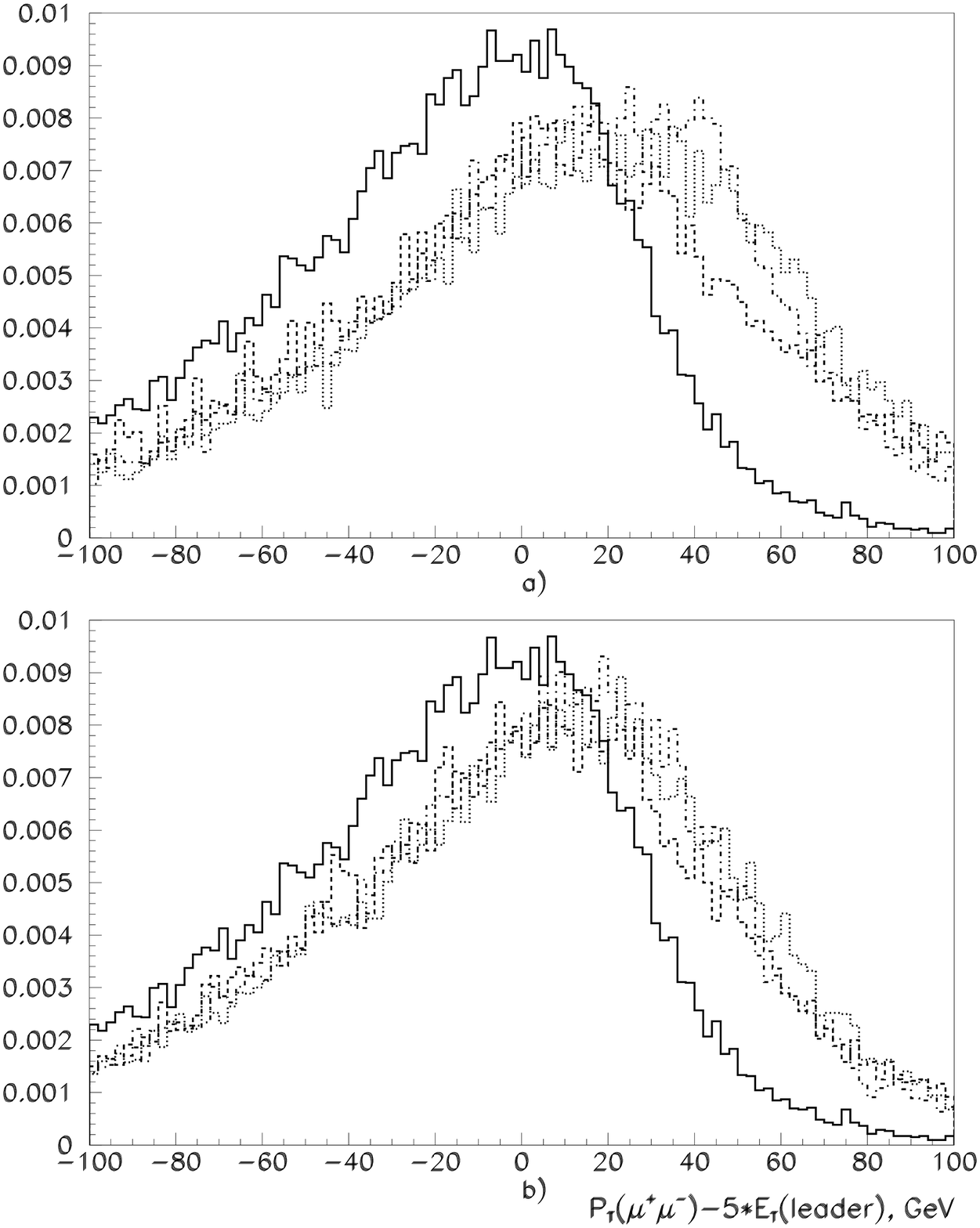, height=180mm}}   
\vskip -5 mm
\caption{The same as in fig.4, but including jet energy ``smearing'',
$\sigma_{E_T}=1.5 \sqrt{E_T}$ GeV.} 
\end{center}
\end{figure}
\end{document}